\documentclass[conference]{IEEEtran}
\IEEEoverridecommandlockouts
\usepackage{cite}
\usepackage{amsmath,amssymb,amsfonts}
\usepackage{algorithmic}
\usepackage{url}
\usepackage{graphicx}
\usepackage{textcomp}
\usepackage{xcolor}
\usepackage{float}

\usepackage{braket}

\usepackage[breaklinks]{hyperref}
\def\BibTeX{{\rm B\kern-.05em{\sc i\kern-.025em b}\kern-.08em
    T\kern-.1667em\lower.7ex\hbox{E}\kern-.125emX}}
\makeatletter
\newcommand{\linebreakand}{%
  \end{@IEEEauthorhalign}
  \hfill\mbox{}\par
  \mbox{}\hfill\begin{@IEEEauthorhalign}
}
\makeatother

\begin{document}


\title{Measuring Sustainability in Multi-Scale High-Performance Computing}

\author{
\IEEEauthorblockN{1\textsuperscript{st} Carlos J. Barrios H.}
\IEEEauthorblockA{\textit{Universidad Industrial de Santander}\\
\textit{SC3UIS, CAGE}\\
Bucaramanga, Colombia \\
\textit{LIG/INRIA Grenoble} \\ 
Grenoble, France\\
\textit{INSA Lyon, INRIA}\\
\textit{CITI Laboratory} \\ 
Villeurbanne, France\\
0000-0002-3227-8651}

\and
\IEEEauthorblockN{2\textsuperscript{nd} Frédéric Le Mouël}
\IEEEauthorblockA{\textit{INSA Lyon, INRIA}\\
\textit{CITI Laboratory}\\
Villeurbanne, France \\
0000-0002-7323-4057}
\and
\IEEEauthorblockN{3\textsuperscript{rd} Yves Denneulin}
\IEEEauthorblockA{\textit{LIG/INRIA Grenoble} \\ 
\textit{Université Grenoble-Alpes, Grenoble-INP}\\
Grenoble, France \\
0000-0002-0340-2094}
}

\maketitle

\begin{abstract}
The transition from traditional High Performance Computing (HPC) to the Computing Continuum emphasizes efficient resource management and sustainable practices across Multi-Scale hybrid architectures. This paper introduces a multidimensional metric framework to characterize these systems and guide deployment strategies for modern workloads. The framework combines Architectural Performance metrics (such as Throughput, Latency, Scalability), System Utilization, and key Sustainability and Accuracy indicators (such as Energy Efficiency and Power Consumption). Using a modular hybrid testbed, experiments reveal complex relationships among metrics, especially the trade-offs between accuracy and energy, and the efficiency of hybrid nodes. The guidelines help identify optimal operating points and lay the groundwork for improving orchestrators and schedulers (e.g., Kubernetes) to assign demanding applications, including AI and Quantum Computing, to suitable system modules, ensuring high performance and sustainability.
\end{abstract}


\section{Introduction}
\label{INT}
Transitioning from High Performance Computing (HPC) to Advanced Computing within the Computing Continuum requires an understanding of how to integrate multi-scale systems. This involves utilizing parallelization and distribution by combining diverse computing resources and multi-scale workloads across multiple systems, ensuring sustainable performance while effectively managing complexity and scalability.

Based on functional requirements, different workloads are allocated among connected computing resources for parallel processing. When viewed horizontally, these resources are equally diverse and integrated into hybrid systems and heterogeneous architectures, such as CPU-GPU based systems. Looking at it vertically, we observe different levels or layers, depending on what we wish to identify; for example, in a continuous computing architecture consisting of various interconnected levels \cite{b1}. The integration and convergence of different computational systems and architectures aim to enhance performance and improve computational efficiency for complex tasks across multiple scales. This methodology, known as hybrid HPC or heterogeneous HPC, leverages the strengths of various computer architectures to handle diverse workloads effectively. To grasp these concepts, one must analyze architectural diversity through the lens of heterogeneity and scaling through the lens of hybridity.

The specialized community has long been developing a comprehensive definition of heterogeneous computing that encompasses the smooth and coordinated deployment of various high-performance machines, including parallel systems, to achieve ultra-fast processing for demanding tasks with diverse computational needs \cite{b2}. In terms of systems, the hybrid computing approach represents the intersection of three broad paradigms for computing infrastructure and use: (1) Owner-centric \emph{(traditional)} HPC; (2) Grid computing (resource sharing); (3) Cloud computing (on-demand resource/service provisioning) \cite{b3}. Definitions based on theoretical assumptions face technical implementations of hybrid computing date back to the sixties, mixing both architectural diversity and the increase in parallel processes and data scale, as noted in \cite{b4}. These paradigms are important because they allow us to identify the attributes necessary to support different use cases that traditional HPC systems must contain, guiding us precisely towards Multi-Scale HPC systems.

\begin{figure}[htbp!]
  \begin{center}
    \includegraphics[width=8.5cm]{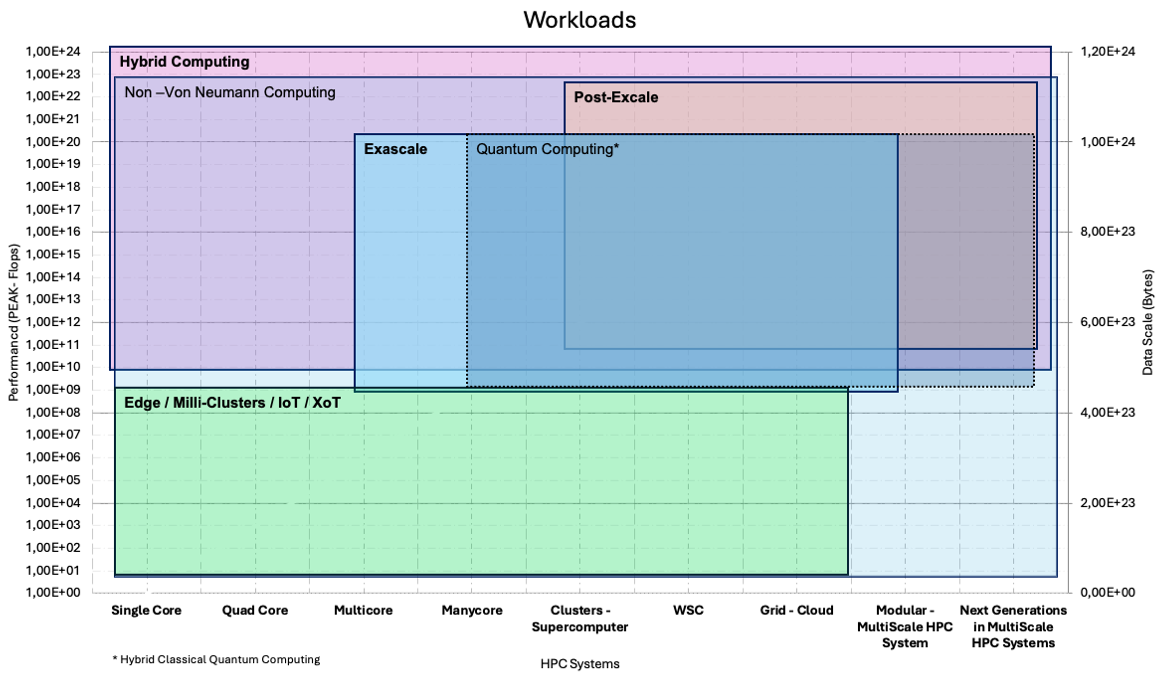}
    \caption{Resources and Workloads Scale}
    \label{fig:WS}
   \end{center}
\end{figure}

Figure \ref{fig:WS} shows the correlation between workloads and scale within the framework of High Performance Computing (HPC). The vertical axes illustrate the performance, specifically the peak performance measured in FLOPS, alongside the data capacity and bandwidth. The classification of large-scale computation depends on the magnitude of operations and the volume of data. Workloads, classified by application type as intensive or massive, are supported by various computer architecture organizations that inherently foster parallelism and High Throughput Computing (HTC).

HPC systems for execution and deployment are tailored to specific workload types, taking into account the technological infrastructure and software platforms used for application development, deployment, and execution. The organization shown in Figure \ref{fig:WS} is directly associated with the processing units in that environment. For instance, the first three categories, single-core, quad-core, and multi-core, reflect a similar architecture that suits monolithic or owner-centric computing. In contrast, many-core and supercomputer clusters scale up, sharing resources and forming interconnected components, underscoring the significance of network capabilities at this scale. This journey begins with Warehouse-Scale Computing (WSC) and extends to Grid, Cloud, and Modular Supercomputing Systems \cite{b5}, all of which imply scalable distributed computing environments. We can characterize computation at scale by examining workloads, HPC systems, performance, and scalability. This includes computing related to Edge, Milliclusters, IoT, or any embedded systems that support parallelism (or \emph{HPC@Pocket} systems \cite{b6}), as well as Exascale and Post-Exascale computing \cite{b7}, hybrid and reconfigurable systems\cite{b8}, and even non-von Neumann architectures that integrate Quantum Computing into these Multi-Scale HPC systems \cite{b9} \cite{b10}. Additionally, it encompasses quantum computing, particularly from a hybrid-computing perspective that integrates classical computer architectures.

Adopting a holistic view provides a thorough understanding of computing systems, marking the shift from traditional HPC to more intricate Multi-Scale environments. This approach facilitates the creation of a performance metric framework. Recognizing the ability to navigate complex, interconnected computational landscapes is essential for adopting a unified approach across various computing paradigms. The heterogeneity and hybrid nature of these systems help clarify behaviors, trade-offs, sustainability, and energy consumption metrics of different features. By exploring expanding scales, applications, and multi-level behaviors, one can outline a Multi-Scale High Performance Computing (Multi-Scale HPC) system that ensures efficient resource allocation and highlights key metrics for system monitoring and design. In other words, we define Multi-Scale HPC as the capability of a system and its software framework to manage workloads across multiple interconnected dimensions of scale: (i) Hardware Scale (from heterogeneous cores to accelerated nodes), (ii) Algorithmic/Computational Scale (from mixed/variable-precision operations to massive multiphysics simulations), and (iii) Temporal/Energy Scale (energy consumption dynamics based on the application phase).

This contribution presents application metrics and multi-level behaviors, aiming to create a metric framework that leverages workload analysis to support monitoring techniques. It also lays the groundwork for designing sustainable Multi-Scale hybrid systems. The discussion covers components for understanding and integrating traditional HPC architectures into distributed, multi-scale systems, as well as guidelines for characterizing these architectures using sustainability and scalability metrics. Addresses diverse workloads for effective deployment, especially in hybrid and heterogeneous architectures.

This paper is structured as follows: after this introduction, Section \ref{SECK} highlights the essential components and characteristics that shape our guidelines. These guidelines are examined in Section \ref{SECR}, where we emphasize the connection between resource utilization and performance, allowing us to identify sustainability and precision metrics. The relevant analysis results are provided in Section \ref{MHPCSC}, followed by a discussion of our further work in Section \ref{DFW}. Finally, Section \ref{CON} concludes our findings.

\section{Key Conceptual Elements and Characteristics}
\label{SECK}

When examining all interconnected components: hardware, software, algorithms, and data, within a framework supported by a Multi-Scale HPC system, it is crucial to effectively integrate and define architectures and environments. Multi-Scale HPC systems require incorporating models that operate at varying scales. For example, in exascale computing environments, Multi-Scale applications must efficiently scale across thousands of nodes, which requires meticulous coordination of computational resources and algorithms. Furthermore, this demands \emph{malleability} \cite{b11} to accommodate diverse workload profiles, ensuring effective utilization of system resources and maximizing performance and throughput.

Integrated architectures in Multi-Scale HPC systems are complex, involving customizable, heterogeneous hardware such as FPGAs, hybrid GPU-CPU systems, and networked clusters. Optimizing these architectures requires a comprehensive approach to components and interactions, managing interdependencies. Balancing computational loads and ensuring efficient data transfer between storage and processing units are vital to improving performance.

Three key conceptual elements can be delineated: first, architectural integration, which presents a methodology for uniting diverse computing paradigms; Secondly, the distributed organization of computing resources, characterized as a distributed HPC system; and finally, Multi-Scale computing, aimed at addressing complexity and scalability \footnote{In effect, Multi-Scale computing and traditional computing differ fundamentally in their approach to problem-solving, particularly in terms of complexity and scalability.} \cite{b12}.

\begin{figure}[htbp!]
  \begin{center}
    \includegraphics[width=8.5cm]{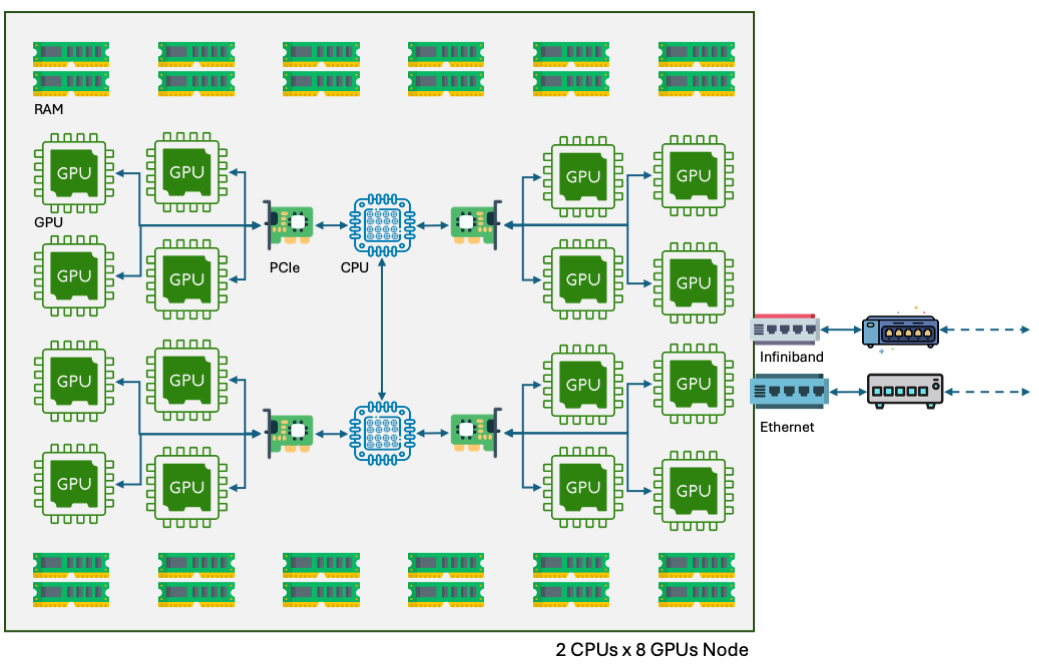}
      \end{center}
  \caption{Organization of a Multi-GPU-CPU HPC Cluster Node}
  \label{fig:NGuane}
\end{figure}

Figure \ref{fig:NGuane} displays the layout of the computational architecture for the servers within a Multi-GPU-CPU HPC cluster node. These nodes are designed to optimize computational performance by leveraging the combined power of CPUs and GPUs. They are essential for addressing complex tasks in AI, scientific research, and data analytics, making them vital components of modern HPC systems. When each node is outfitted with this hybrid setup, the system can effectively handle increasing workload complexity and data volume. For instance, applications that use Massive Parallel Processing (MPP) benefit from this configuration, which requires multilevel parallelism to achieve scalability.  

\begin{figure}[htbp!]
  \begin{center}
    \includegraphics[width=8.5cm]{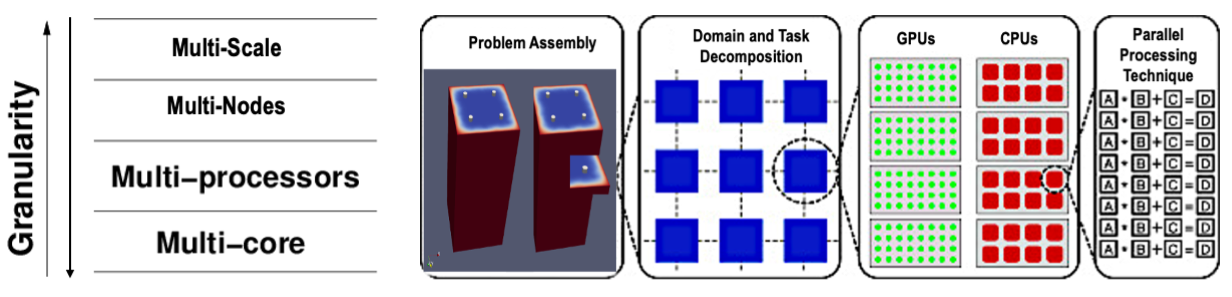}
      \end{center}
  \caption{Multilevel Parallelism}
  \label{fig:MultiLevelP}
\end{figure}

Multilevel parallelism is the ability to exploit parallelism at multiple levels within a computational task. This concept is especially valuable for improving the efficiency and speed of computations by leveraging different forms of parallelism, including coarse-grained, fine-grained, and instruction-level parallelism. In a complex scientific application, fine-grained computing is directed to the processing cores, while coarse-grained components are allocated to processors or nodes. This allocation increases committed infrastructure resources based on the specific computation requirements. Additionally, this observed behavior improves scalability, allowing Multi-Scale HPC systems to accommodate parallelism as needed and support more complex applications effectively. Figure \ref{fig:MultiLevelP} illustrates a well-known pipeline for exploiting multilevel parallelism, noting that varying granularity introduces different architectural elements.

This principle has been recognized and utilized for some time in various scale-up workloads, known as ultrascale systems \cite{b13}. Nevertheless, challenges related to resource allocation and efficient execution persist, especially due to the diversity of required workloads and resources; for instance, the need for effective communication strategies, job scheduling techniques \cite{b13a}, and optimization strategies \cite{b14}.

\begin{figure}[htbp!]
  \begin{center}
    \includegraphics[width=8.5cm]{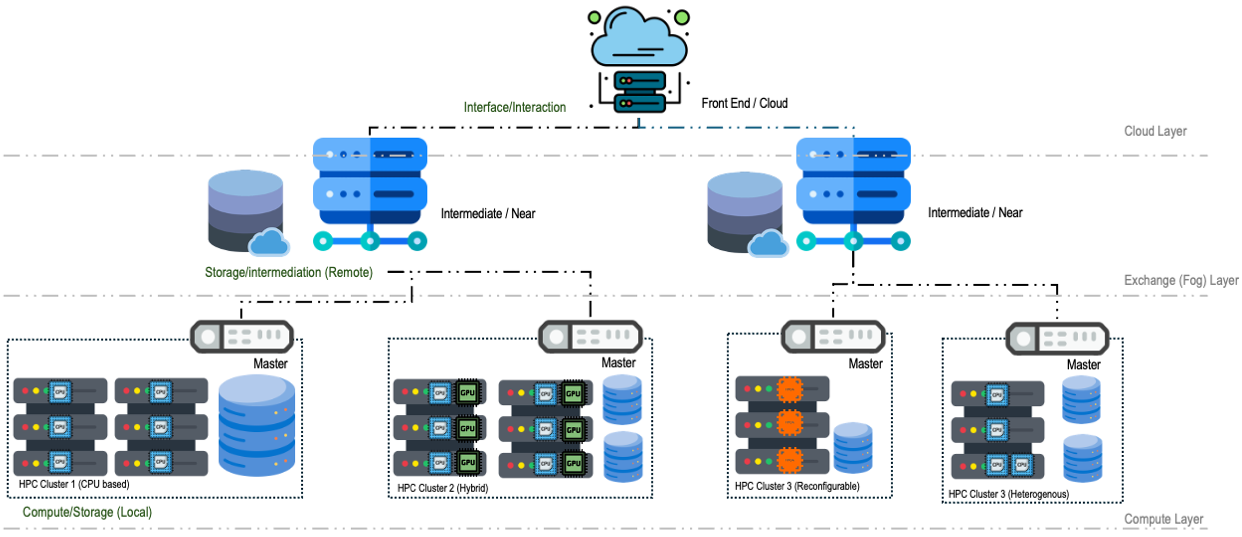}
      \end{center}
  \caption{Multilevel and Multi-Layer HPC Architecture}
  \label{fig:MultiLevelA}
\end{figure}

Technically, it is possible to observe a Multi-Scale HPC system that defines architectural layers, identifies roles and interactions according to functionality, and uses levels as an abstraction to describe the hierarchy within the multi-tier architecture. Figure \ref{fig:MultiLevelA} shows an abstraction that identifies different components of a Multi-Scale HPC system.  

A Multi-Scale HPC system can comprise various types of platforms. Figure \ref{fig:MultiLevelA} shows four categories of infrastructure. Horizontally, these include high-performance pure CPUs, hybrid CPU-GPU systems, FPGA-based reconfigurable platforms, heterogeneous systems with varying processing units per node, and small pure CPU architectures. These emphasize both local computing and storage. Vertically, each layer represents a specific functional abstraction, starting from computing in the core layer, moving through the fog/exchange layer, which handles intermediate resources as scale and complexity increase, and finally reaching the interaction/interface layer at the cloud level.

These two dimensions integrate architectural designs and HPC configurations into distributed Multi-Scale (HPC) systems by analyzing feature overlap across vertically and horizontally defined parameters and metrics. For instance, when examining horizontal factors, it is important to consider the number of processing cores or accelerators, the level of process support, and the grain types supported. Vertically, one must assess the algorithm's complexity, the tasks or jobs to be performed, and scalability. This method promotes alignment between scalability, performance, interoperability, and resource management, aiding in the development of workload profiles.

This paper does not define I/O performance, as it was held constant to isolate the relationship between compute precision and energy. However, tools that use alternative methods can help in understanding complex workloads. Evaluating these workloads improves characterization, enabling storage systems to optimize I/O performance for specific HPC tasks. This approach also applies to modular and Multi-Scale HPC systems \cite{b15}.

\section{Multidimensional Metric Framework}
\label{SECR}

Key metrics are vital for evaluating HPC system performance, especially as HPC moves toward Multi-Scale distributed systems. They assess system performance and resource usage, support performance evaluation through holistic, monitoring-based analysis and visualization tools \cite{b16}, optimize current systems, and guide future architectural design for complex Multi-Scale computations. We identify two main metric categories: architectural capabilities and environmental behavior, with additional metrics, such as energy efficiency, at their intersection. We propose guidelines for applying these metrics to ensure smooth deployment across hybrid architectures and multilevel functions.

Rather than evaluating standard isolated metrics (e.g., peak FLOPS, raw throughput, or network latency) that fail to capture energy and precision tradeoffs in heterogeneous environments, we formulate composite operational indicators.  We map execution attributes into a multi-scale vector:

\begin{equation}
\mathbf{\Theta} = \langle EERC, CF, RA \rangle
\label{eq:e0}
\end{equation}

Where: \textbf{Energy-Efficiency Resource Conversion ($EERC$):} Measures effective computational output per watt-hour, considering power caps. \textbf{Algorithmic Complexity Factor ($CF$):} Evaluates runtime scaling based on overhead $\mathcal{O}(f(n))$, especially under mixed-precision regimes, and \textbf{Required Accuracy ($RA$):} Penalizes overkill by scaling energy efficiency against error tolerance limits. These metrics influence the Characteristic Ratio ($CR$) in Equation \ref{eq:eq1}, going beyond static benchmarks.

\subsection{Metric Framework Model}

\begin{figure}[htbp!]
\begin{center}
\includegraphics[width=8.5cm]{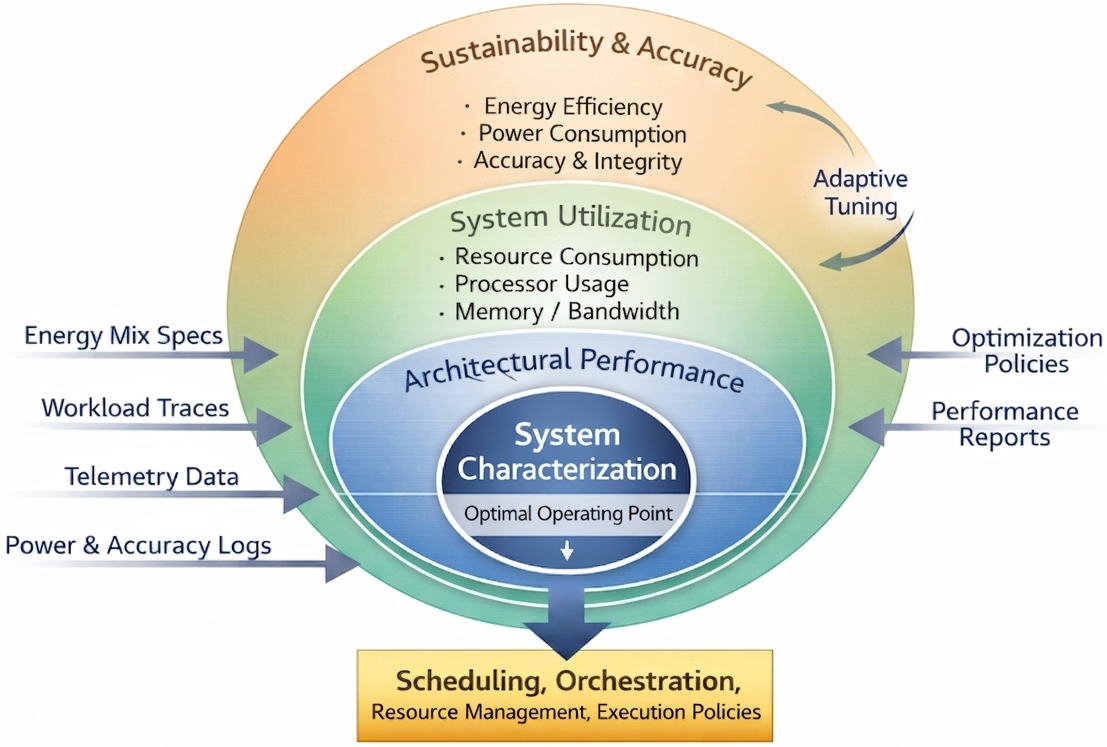}
\end{center}
\caption{Multi-Dimensional HPC Metric Framework}
\label{fig:CM_MF}
\end{figure}

Before displaying performance metrics, users can define the categories in the visual model depicted in Figure \ref{fig:CM_MF}. This model serves as a conceptual framework for understanding how critical metric categories converge and aids in system characterization.

Figure \ref{fig:CM_MF} emphasizes that evaluating systems requires moving beyond individual measures to an analysis of their convergence:
\begin{enumerate}
    \item Architectural Performance (e.g., Throughput, Scalability)  focuses on the system's raw capability and speed.
    \item System Utilization (e.g., Processor, Memory, Network Bandwidth)  measures how efficiently internal resources are being used.
    \item Sustainability and Accuracy (e.g., Energy Efficiency, Accuracy) capture the integrity of the results and the system's environmental cost.
\end{enumerate}

The central convergence element characterizes the optimal system by linking metrics, such as the Energy Efficiency Ratio, derived from performance and power data. System characterization uses operational data to identify the optimal operating point, guiding scheduling to improve hardware efficiency and performance.

\subsection{Architectural Performance Metrics}
Our holistic approach assesses architectural performance metrics like latency, throughput, and memory bandwidth to improve system performance. These metrics help in designing efficient, scalable, and reliable systems. We propose using the following metrics :

\begin{itemize}
\item Throughput: This metric quantifies the volume of workload supported as efficient processing capabilities.
\item Latency: The time required to complete a single operation or task.
\item Scalability: This metric measures a system's ability to handle increased workloads by adding resources. It includes linear scalability, where performance increases proportionally, and sub-linear scalability, where gains lessen with more resources. Understanding this helps develop strategies like strengthening large-scale applications within the broader HPC and data analytics convergence in Multi-Scale HPC architectures \cite{b17}.
\end{itemize}

Latency and throughput are key system performance indicators. Low latency ensures quick responses, while high throughput enables handling multiple requests simultaneously. Workloads impact both; increasing workloads require scalable systems that adapt without performance loss. Scalability is vital for robust design.

\subsection{System Utilization Metrics}
The utilization metrics show the impact of the resource. System metrics reveal how efficiently resources are used, identify bottlenecks, optimize performance, and prevent under or overuse. The following are the key metrics of this work.

\begin{itemize}
\item Processor Utilization: This shows the percentage of Processor capacity used during computations. High utilization indicates efficient resource use.
\item Memory Usage: Monitoring memory usage helps ensure applications do not exceed available memory, which can lead to performance degradation or failures.
\item Network Bandwidth: In distributed systems, the volume of data transmitted over a given time frame is crucial. High bandwidth ensures effective data exchange among nodes.
\end{itemize}

As with architectural metrics, there are interdependencies among performance indicators. For instance, high Processor Utilization often coincides with elevated memory usage in memory-intensive applications. However, if memory bandwidth reaches its limit, the Processor Unit may be underutilized while waiting for data. Similarly, Processor Unit utilization can increase in network-intensive applications due to packet processing overhead. Monitoring these metrics together reveals inefficiencies. Applications that transfer large amounts of data over the network also require significant memory for buffering and caching \cite{b18}.

\subsection{Sustainability and Accuracy Metrics}

Sustainability and Accuracy metrics for Multi-Scale HPC systems are used to measure and optimize resource efficiency, energy consumption, and overall sustainability, particularly to support HPC systems for Artificial Intelligence and other massive, intensive workloads.

\begin{itemize}
\item Energy Efficiency: This metric evaluates the total energy consumed to complete a specific computational task. It is particularly useful for comparing the efficiency of different HPC systems, application executions, or configurations.

\item Power Consumption: As energy efficiency becomes increasingly important, measuring HPC systems' power consumption helps assess their environmental impact and operational costs.

\item Accuracy: Ensuring that the results produced by the system meet the required precision and correctness standards is vital, especially in scientific and Artificial Intelligence (AI) or Deep Learning (DL) computations.
\end{itemize}

This research emphasizes the importance of refining workload-specific accuracy standards, as different Multi-Scale HPC tasks require varying levels of precision. Scientific simulations, for instance, demand extremely high precision, such as double-precision floating-point, to prevent errors and ensure valid results. In contrast, AI and deep learning often use lower-precision formats, such as $FP16$ or lower, to improve performance and reduce energy use with little impact on accuracy.  Quantum computing precision hinges on maintaining coherence and minimizing errors for reliable calculations. Additionally, a thorough, multi-level analysis is essential, assessing accuracy at various stages: instruction-level, core-level, node-level, and inter-node communication, rather than only at the final output. While current evaluations focus on the minimal impact on heterogeneous elements of a Multi-Scale system, it is also vital to recognize that convergence depends on other metrics; a system should be characterized not only by speed or efficiency, but also by the accuracy it provides at those levels. For example, the Optimal Operation Point \emph{(OOP)} \cite{b18a} can be defined by a function $OOP=f(Throughput, Scalability, EE, Accuracy)$, where $EE$ represents energy efficiency. Incorporating these factors allows precision to serve as a comprehensive metric for complex and diverse workloads.

Characterizing the performance and sustainability of Multi-Scale HPC systems is important for ensuring the accuracy and energy efficiency of artificial intelligence applications. Consequently, all metrics must be harmonized within this multi-level framework, emphasizing the convergence of measurements to define and delineate these systems. However, it is important to formalize the different metrics and associate them with quantifiable ratios.

\subsection{Characteristic Ratio}
Taking into account the different dimensions of the metrics, it is possible to introduce a formalized structure for calculating and weighting the characteristics of the system, defining a \emph{Characteristic Ratio (CR)} that facilitates the transition from guidelines to a framework. Then, we propose the use of the formula:

\begin{equation}
\text{CR} = \frac{\mathcal{A} \cdot \text{EERC}}{\text{CF} \cdot \text{RA}}\label{eq:eq1}
\end{equation}

Where $\mathcal{A} = \sum_{i} w_i M_i$ represents the composite Architectural Performance index derived from normalized throughput, latency, and scalability; $\text{EERC}$ is the Energy-Efficiency Resource Conversion ratio; $\text{CF}$ denotes the algorithmic complexity factor; and $\text{RA}$ specifies the workload's required accuracy.

It is important to note that \emph{EERC} as Energy Efficiency Ratio is explicitly defined as $\frac{Throughput}{Power}$, and taking into account that the Complexity Factor \emph{(CF)} is a variable based on the difficulty or interdependence of the workload in execution, a dynamic, dimensionless measure reflecting the computational difficulty and interdependency within a workload. For this approach, $CF$ is assigned based on algorithmic complexity: $O(n)$ workloads are assigned a CF of $1.0$ while $O(n^2)$ or higher interdependencies are assigned a CF of $1.5$. Additionally, the Required Accuracy \emph{RA} must specify the level of numerical precision and correctness that the workload demands of the system. This is useful for balancing performance and sustainability. Evidently, the RA in the denominator of the equation \ref{eq:eq1} penalizes when the workload demands a very high level of accuracy. This approach to applying convergence across multidimensional elements is often used for Multi-Scale (and large-scale) systems because simple maximum-speed metrics cannot effectively capture the 'extreme heterogeneity.' Therefore, its numerator should be the sum of architectural metrics ($\sum Architectural Metrics$), and the metric should focus on the system's internal features rather than on raw performance levels alone.

\section{Multi-Scale HPC Systems Characterization}
\label{MHPCSC}

According to our modular architecture guidelines for a Multi-Scale HPC system, the three key stages help us understand architecture, performance metrics related to system utilization and sustainability, and accuracy metrics for managing workloads. Architectural complexity arises from heterogeneous computing resources, primarily driving horizontal scaling that enhances support by adding more nodes. However, as shown in the proposed testbed, four distinct processing capabilities and configurations influence both individual and module performance.

As mentioned previously, various infrastructure configurations enable observation from elevated platforms. For instance, some platforms rely solely on CPUs with fixed core counts and processor frequencies, while others integrate CPU-GPU combinations with different core counts and GPU families on interconnected nodes. The data collected comes from four modules: the first module includes nodes 1 to 6 with low-profile CPU-only configurations; the second module, covering nodes 7 to 20, employs a CPU-GPU setup with four mid-range CPUs and four mid-range GPUs; the third module consists of nodes 21 to 28, which include two high-range CPUs and two high-range GPUs; and the final module, containing nodes 29 and 30, features a different family of two high-range CPUs and two high-range GPUs per node. This configuration is available across various frameworks. For example, in extensive systems like Grid5000\footnote{\url{https://www.grid5000.fr}} or hybrid platforms presented in the SCALAC system\footnote{\url{https://scalac.redclara.net}}, we can identify distinct modules: a basic configuration with only a CPU module comprising two Intel Xeon CPUs operating at 2.4 GHz with 24 cores, providing 102 GB of RAM. Another set of CPU-GPU nodes features Intel Xeon CPUs running at 2.67 GHz (16 cores) and another at 2.4 GHz (64 cores), along with 2 NVIDIA K20 and K80 GPUs and 320 GB of RAM. Furthermore, CPU-GPU nodes employ AMD EPYC 9534 and 9554 processors operating at 2.5 GHz, paired with 2 AMD Instinct MI210. All modules are interconnected through high-bandwidth Gigabit Ethernet and InfiniBand networks simultaneously.

\begin{figure}[htbp!]
\begin{center}
\includegraphics[width=8.5cm]{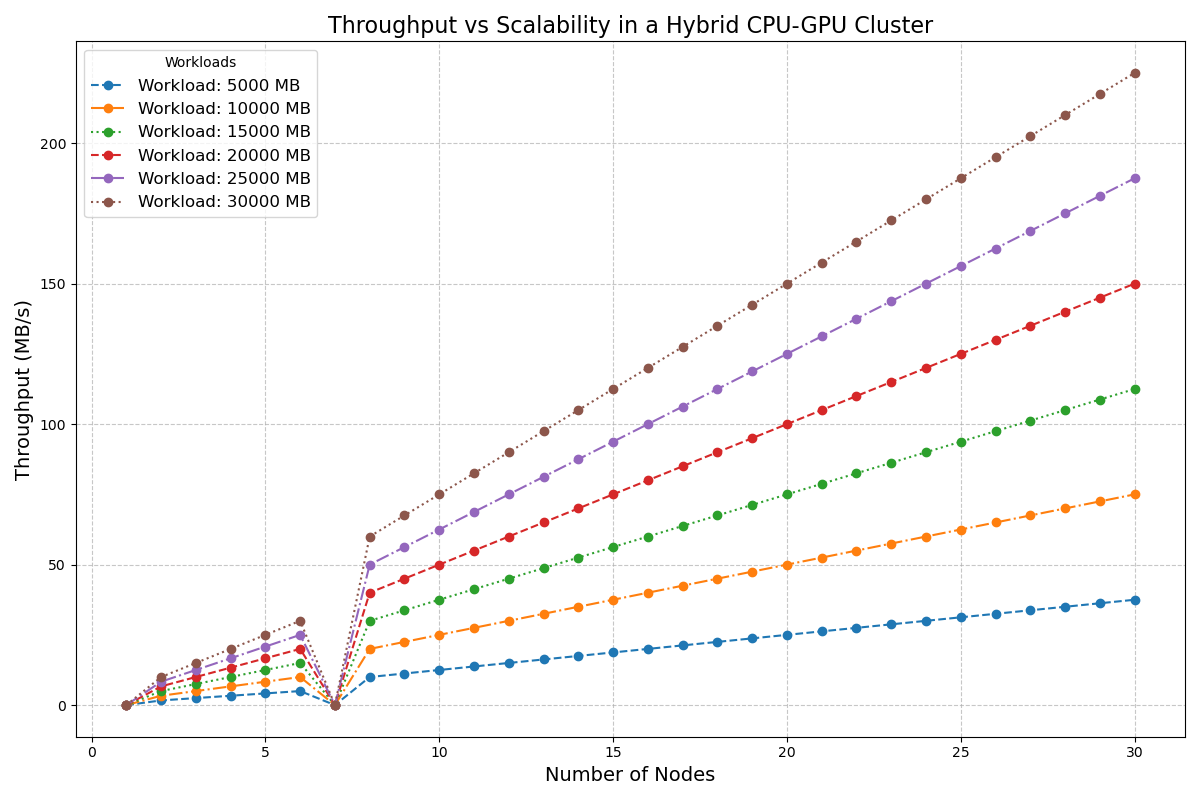}
\end{center}
\caption{Throughput}
\label{fig:ThrSCABi}
\end{figure}

The evaluation of workloads ranging from $5000MB$ to $30000MB$ is conducted across a spectrum of 2 to 30 diverse nodes, achieving average processor utilization between 89.7 and 97.3 percent. Throughput is determined by dividing the workload by the total time, adjusting for the number of nodes, and then refining it using a logarithmic efficiency factor \cite{b19}. Figure \ref{fig:ThrSCABi} illustrates the relationship between throughput and scalability, detailing how throughput varies with the number of nodes within the hybrid CPU-GPU cluster.

This expected behavior is observed across different workloads with high memory usage, even on similar platforms. Various factors can lead to high memory utilization, including process types, memory management strategies, and system architecture. However, in our tests, it is mainly due to the process itself. The system's heterogeneous and hybrid architecture helps manage these trade-offs, allowing for resource optimization. For example, tasks requiring low-to-moderate accuracy can be assigned to energy-efficient components such as GPUs or specialized low-power cores. In contrast, high-precision tasks are reserved for traditional CPU nodes, accepting increased energy use to maintain accuracy.

\begin{figure}[htbp!]
  \begin{center}
    \includegraphics[width=8.5cm]{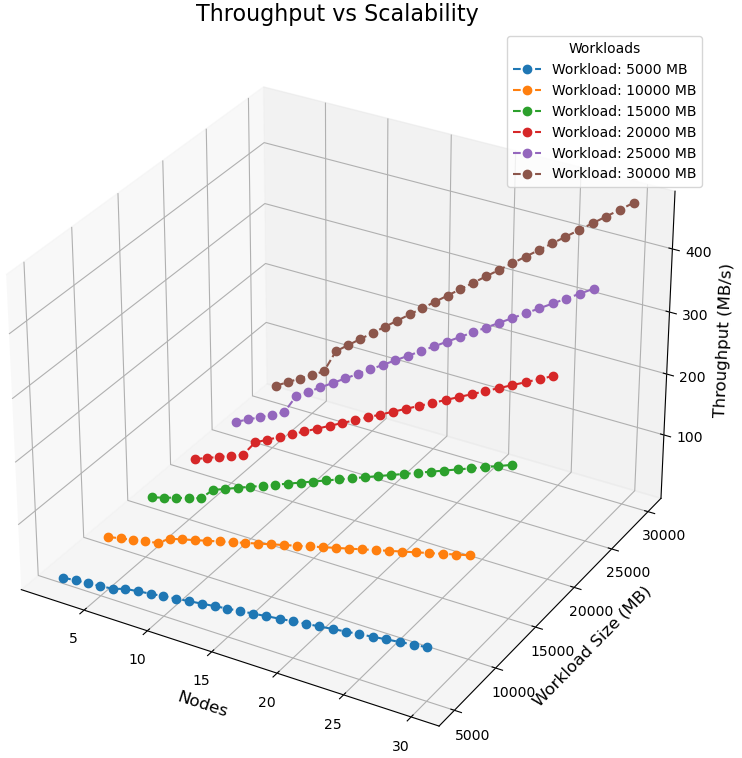}
      \end{center}
  \caption{Throughput and Scalability}
  \label{fig:ThrSCA}
\end{figure}

However, following our established guidelines,  it is crucial to examine the relationships among throughput, scalability, and the number of nodes while considering the deployed workloads, as illustrated in Figure \ref{fig:ThrSCA}.  

\begin{figure}[htbp!]
  \begin{center}
    \includegraphics[width=8.5cm]{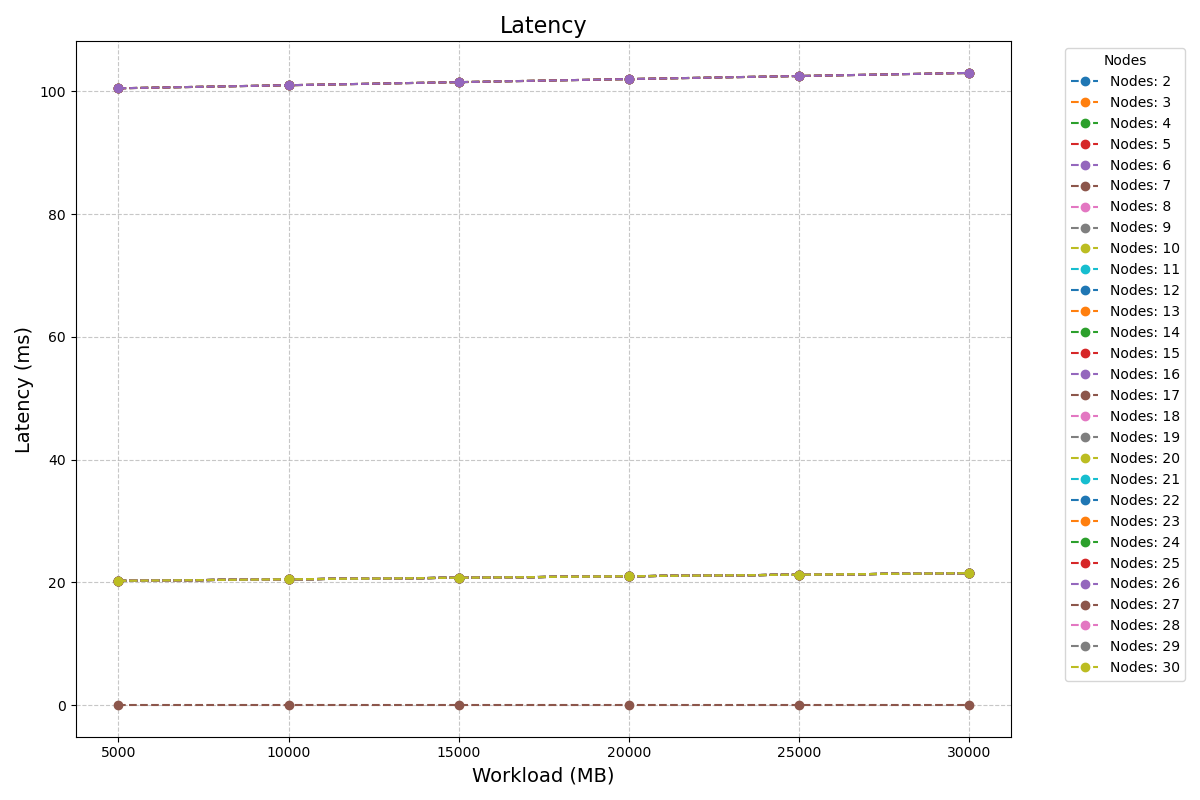}
      \end{center}
  \caption{Latency}
  \label{fig:Lat}
\end{figure}

Throughput and scalability are key metrics in computing, particularly in the realm of Multi-Scale HPC. High throughput facilitates efficient handling of complex computations, while scalability ensures that systems can expand and adapt to growing demands without compromising performance. Together, these factors are essential for enhancing computational capabilities. Figure \ref{fig:ThrSCA} presents the optimal workload that maximizes throughput given the platform's characteristics. However, there is significant overloading for workloads ranging from 20000 to 30000 MB, as throughput is limited by the slowest component, particularly noticeable during transitions between modules 1 and 2 (only CPU and CPU-GPU modules). In contrast, performance remains stable when shifting from module 3 to module 4 (CPU-GPUs Module but with two different processors and GPU families). By analyzing this behavior, insights can be gained regarding the system's capacity to manage increasing workloads effectively. At the same time, latency varies with workload size; specifically, high latency is observed on nodes 1 to 6, while low latency is observed in the range of 8 to 30 nodes, with a slight increase as workloads grow. Consequently, due to the architectural configuration, high bandwidth is provided to modules with the most resources, as shown in Figure \ref{fig:Lat}.

\begin{figure}[htbp!]
  \begin{center}
    \includegraphics[width=8.5cm]{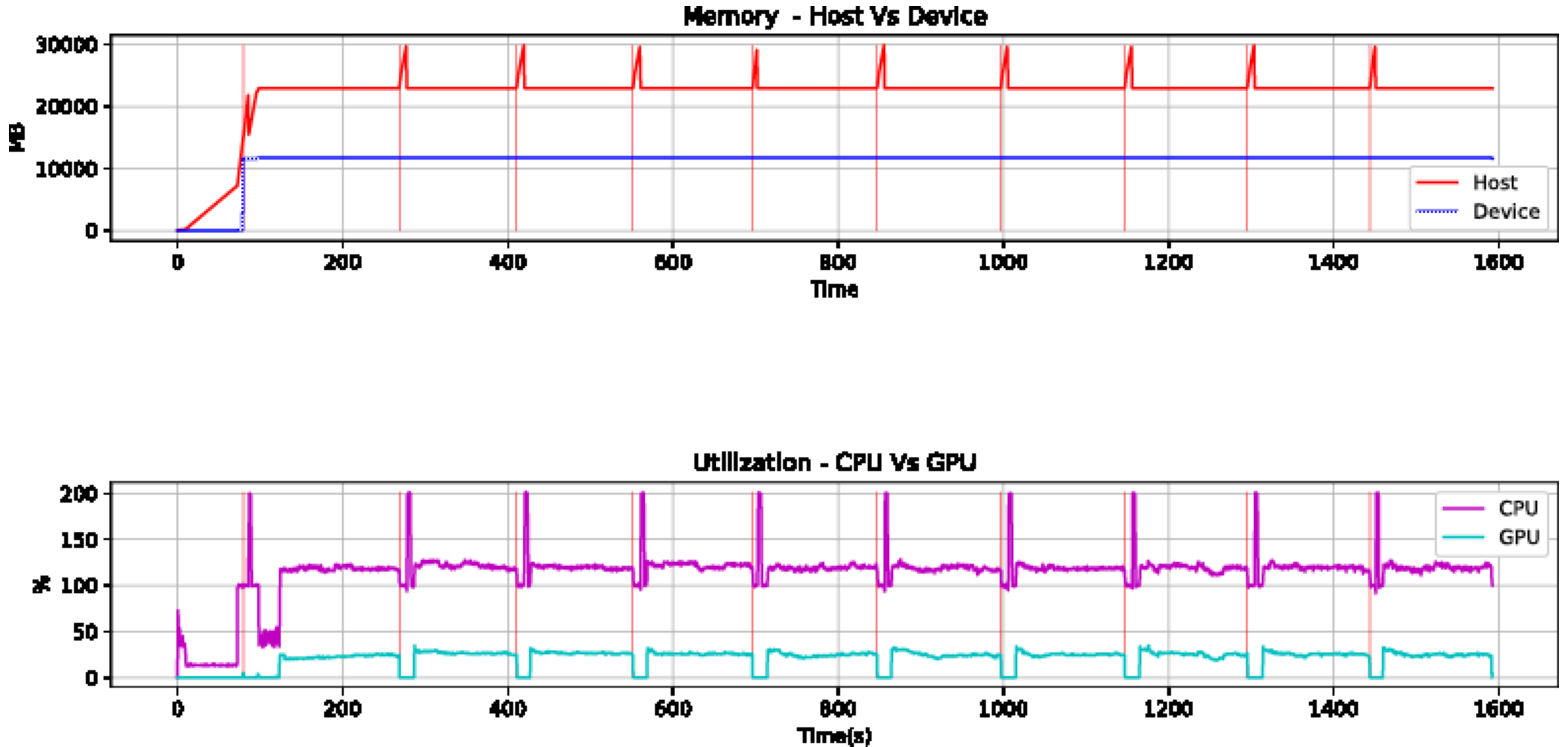}
      \end{center}
  \caption{Architecture Performance}
  \label{fig:AP}
\end{figure}

Figure \ref{fig:AP} illustrates the memory consumption of both the CPUs and the GPUs, encompassing elements of both the host and the device throughout the trials carried out. A clear depiction emphasizes the efficiency and interaction between the host and device, highlighting the potential architectural variability that may arise from inefficient interactions between accelerators. The involvement of the CPU and GPU demonstrates the impacts on these resources when diverse modules participate in the overall execution.

\begin{figure}[htbp!]
  \begin{center}
    \includegraphics[width=8.5cm]{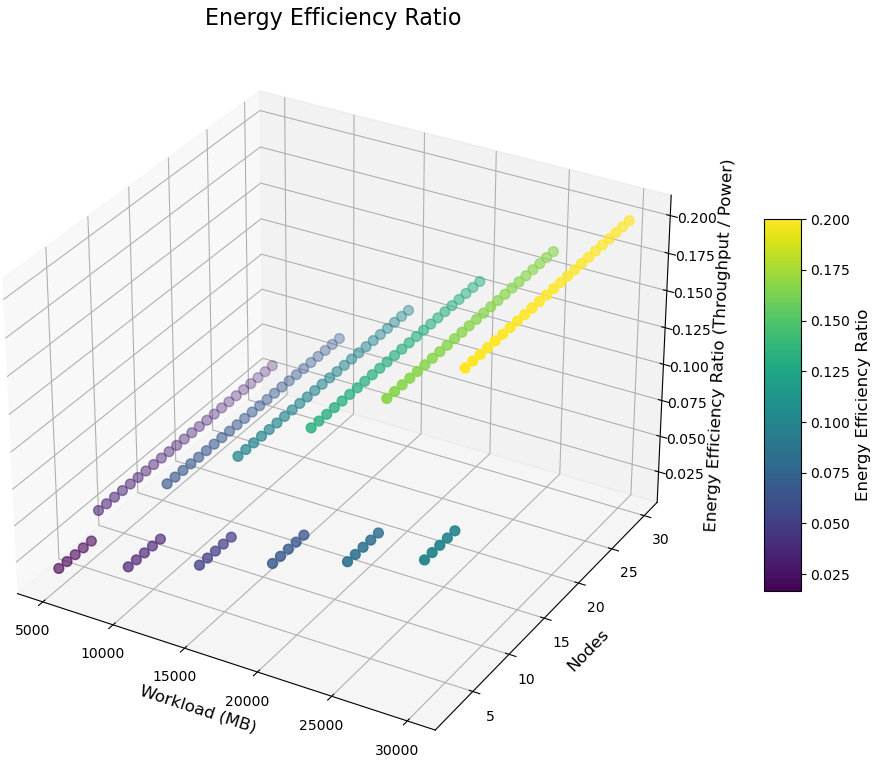}
      \end{center}
  \caption{Energy Efficiency Ratio}
  \label{fig:EnergyEfficencyRatio}
\end{figure}

When assessing architectural features, CPU nodes typically have higher power consumption, even though they run at reduced frequencies. In contrast, CPU-GPU nodes demonstrate superior energy efficiency, averaging around 100 Watts per node. The energy efficiency ratio is determined by comparing the throughput to the overall power consumption. We examined this correlation by analyzing the total energy used to execute all tasks on the active nodes. The differences illustrated in Figure \ref{fig:EnergyEfficencyRatio} stem from variations in power consumption among the modules, with CPU-GPU nodes reflecting the aggregate of both types. When combined within CPU-GPU nodes, their total power usage accurately represents the sum of their individual inputs, significantly affecting energy efficiency, performance, and sustainability. Then, by leveraging the strengths of each module and optimizing their combined usage in CPU-GPU nodes, it is possible to achieve a balance between performance and energy efficiency, which is increasingly important for Multi-Scale HPC systems as will be discussed in Section \ref{DFW}.

In this paper, we introduce the relationship between accuracy and energy consumption, enabling the use of different analysis mechanisms, such as trade-off curve analysis and architectural influence analysis. In Multi-Scale hybrid HPC architectures, balancing precision and energy consumption is crucial because these systems combine quantum, analog, and classical components. Higher accuracy usually consumes more energy, which is challenging in low-power settings. Techniques such as reducing precision or selective processing can lower energy consumption while supporting tasks like machine learning inference or real-time analytics. This trade-off is complex in hybrid systems where subsystems differ in speed, accuracy, and energy needs. Managing this balance is essential for optimal performance and sustainability without sacrificing key results.

\begin{figure}[htbp!]
  \begin{center}
    \includegraphics[width=8.5cm]{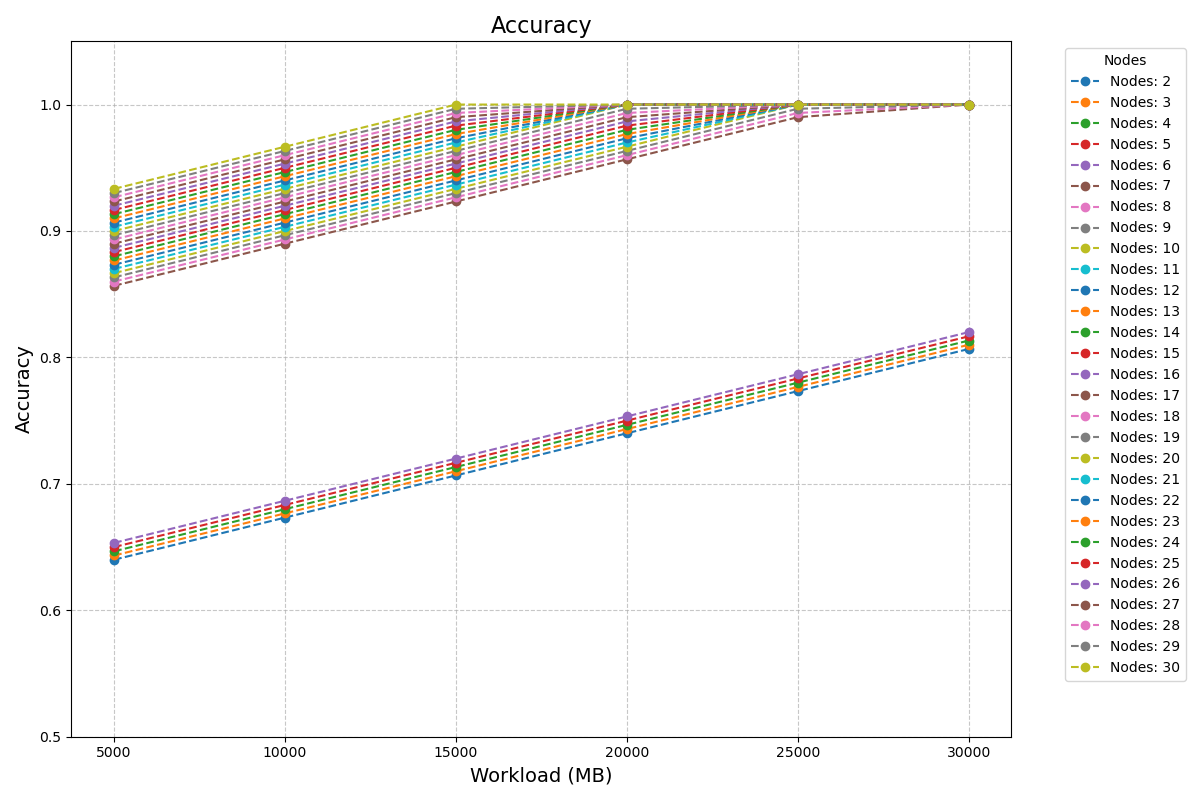}
      \end{center}
  \caption{Accuracy}
  \label{fig:Accuracy}
\end{figure}

Figure \ref{fig:EnergyEfficencyRatio} is depicted in three dimensions to illustrate the system and incorporate the metrics discussed above. While efficiency improves with increased processing support, energy consumption also increases as more nodes are added. Additionally, understanding the relationship between accuracy and energy is vital for a complete analysis. To facilitate this, we used a simple backpropagation algorithm \cite{b20}, which can be characterized in terms of matrix multiplication. While backpropagation serves as a proxy for the matrix-heavy operations used in AI, subsequent validation with the MLPerf benchmark suite is planned to further demonstrate the framework's versatility across diverse production workloads. The initial accuracy is depicted in Figure \ref{fig:Accuracy}.

\begin{figure}[htbp!]
  \begin{center}
    \includegraphics[width=8.5cm]{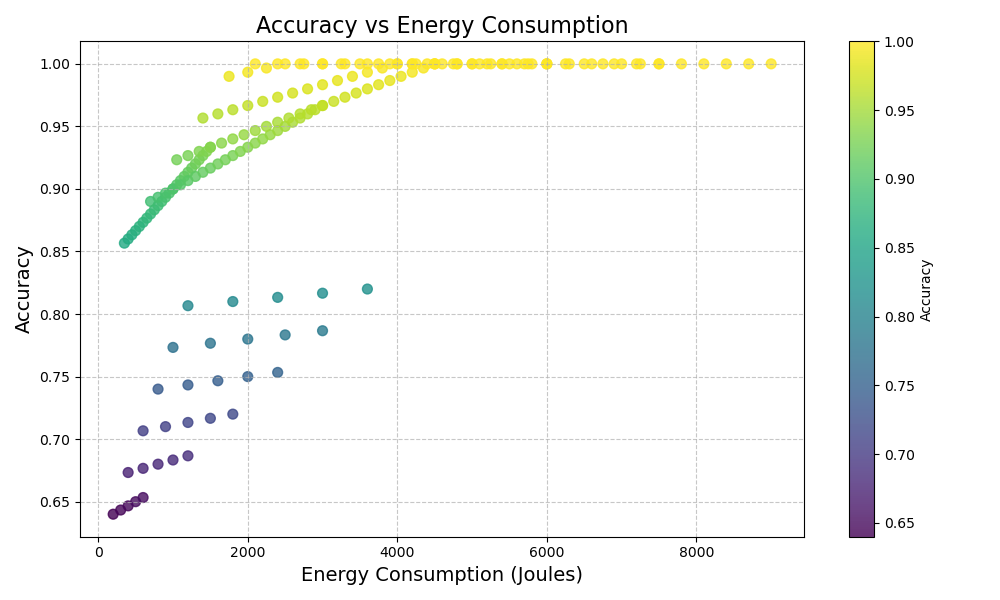}
      \end{center}
  \caption{Accuracy vs Energy Consumption}
  \label{fig:EnergyAccuracy}
\end{figure}

More accuracy is achieved with increasing workloads and node counts, and GPU nodes deliver better performance. By analyzing the experience, we find that both accuracy and energy consumption depend on workload size and node configurations. Energy consumption increases with workload size and total node power, as illustrated in the figure. In Figure \ref{fig:EnergyAccuracy}, energy consumption is measured in joules, and the accuracy is limited to a maximum value of 1.0. 

The relationship between Accuracy and Energy Consumption motivates introducing a \emph{Required Accuracy (RA)}, a performance indicator that is part of the \emph{CR} in formula \ref{eq:eq1}. Unlike the Energy Delay Product (EDP) \cite{b20EDP}, which focuses on time-energy trade-offs, our CR metric offers a more nuanced approach to resource selection by penalizing \emph{Accuracy Overkill} in simple tasks. Accuracy standards depend on hardware features, but this paper relies on manufacturer data without in-depth analysis. I/O performance, which is significant in HPC workloads, is not included but will be in future CR versions, with an I/O Wait-Time coefficient. I/O bottlenecks are discussed in upcoming research. The convergence analysis in Figure \ref{fig:EnergyAccuracy} highlights the trade-off between quality and environmental cost, supporting the inclusion of $\text{RA}$ in the denominator of $\text{CR}$. Data show saturation \cite{b20b}, where small accuracy reductions (e.g., $98\%$ vs. $99.5\%$) significantly cut energy ($25\%$) with minimal quality loss, but higher fidelity costs more energy for little gain. Setting an optimal $\text{RA}$ based on this trade-off helps $\text{CR}$ penalize over-resourced accuracy, promoting sustainability without sacrificing quality. Results show that effective characterization indicates potential support for Multi-Scale HPC systems and helps inform deployment policies and the development of new metrics.

\subsection{Extending the \texorpdfstring{$CR$}{CR} Metric to Quantum and Edge Domains}
To avoid overgeneralization, our current experimental validation focuses on classical heterogeneous HPC nodes. However, the mathematical formulation of the Characteristic Ratio ($CR$) is designed to accommodate Quantum and Edge paradigms by mapping their technology-specific parameters into our multidimensional vector $\mathbf{\Theta} = \langle EERC, CF, RA \rangle$:

\begin{itemize}
   \item \textbf{Quantum Computing Formulation:} For quantum-classical hybrid execution, the accuracy parameter $RA$ is redefined as a function of Quantum Gate Fidelity ($\mathcal{F}$) and coherence time limits ($T_1, T_2$), such that $RA_Q = f(\mathcal{F}, T_1)$. The algorithmic complexity $CF$ incorporates the circuit depth ($D$) and the qubit width ($Q$). The energy efficiency ratio $EERC$ accounts for the total cryogenic cooling overhead per executed circuit.
   
    \item \textbf{Edge/IoT Constrained Systems:} In edge environments, $EERC$ is adjusted to penalize network transmission overhead ($\Delta t_{net}$) alongside computational power consumption ($P_{edge}$), reflecting tight energy budgets and bandwidth constraints:

\begin{equation}
\text{EERC}_{\text{edge}} = \frac{\text{Throughput}}{P_{\text{edge}} + \beta \cdot \text{Latency}_{\text{net}}}
\label{eq:eerc_edge}
\end{equation}

This formulation describes $\text{EERC}_{\text{edge}}$, a refined metric of energy efficiency tailored for distributed edge environments. The numerator reflects the effective computational output, while the denominator expands local power dissipation ($P_{\text{edge}}$) to include network communication overhead ($\text{Latency}_{\text{net}}$). The parameter $\beta$ serves as an application-specific penalty coefficient that balances offload latency and transmission energy costs, preventing the framework from favoring nodes with limited energy or high delays.

\end{itemize}

Extending the CR formulation to quantum and edge regimes confirms its viability as a generalizable metric and provides the necessary theoretical foundation for addressing sustainability and deployment trade-offs.

\section{Discussion and Further Work}
\label{DFW}

In high-workload Multi-Scale HPC systems, especially those engaged in intensive processing such as AI and Quantum Computing, monitoring sustainability metrics requires a hybrid strategy to ensure optimal deployment. Surpassing energy efficiency thresholds can reduce effectiveness, due to physical limitations \cite{b21}, and limited performance benefits of higher energy use. This raises questions about practicality, environmental responsibility, and cost. These systems aim to provide tailored capabilities within each module by deploying workloads to improve performance. For example, in CPU-and-GPU-combined nodes, power consumption depends on the workload and module efficiency. Integrated GPUs that share memory with CPUs can reduce power and heat, making them suitable for energy-constrained settings.

Naturally, accuracy and correctness standards must be taken into account. Hardware components, such as support for various floating-point formats, algorithms like error-correction codes, and the application's tolerance for approximation, all affect the system. However, this document primarily relies on manufacturers' standard values and metrics without delving into them. Additionally, although I/O performance is crucial for understanding complex workloads and optimizing storage systems in HPC environments, it is not explicitly defined here. Future iterations of the CR formula will integrate an I/O Wait-Time coefficient to account for data-intensive bottlenecks. I/O often acts as a significant bottleneck in many modern HPC and data-intensive applications. These aspects are explored in further research initiatives.

To implement monitoring systems that analyze metrics and facilitate analysis, especially considering heterogeneity, a factor w can be introduced into equation \ref{eq:eq1}. This factor, known as the weighting factor $\mathbf{w_{m,W}}$ \cite{b21b}, transforms generic hardware performance data into a workload-aware metric. It emphasizes architectural features critical to the specific job, such as assigning a higher weight to Memory Bandwidth for data-intensive workloads. However, applying regression coefficients (weights) to low-level hardware counters (architectural metrics) to predict a higher-level composite outcome (e.g., performance or energy) is a standard practice in performance and power modeling and could make the model more complex or introduce biases in the correlations within a holistic approach.

The evolution of applications, along with the complexity and behavior of workloads in Multi-Scale HPC systems, requires new paradigms and guidelines for analyzing and evaluating performance and sustainability that extend beyond traditional metrics. These new frameworks should consider energy efficiency, workload-specific performance, scalability, resilience, carbon footprint, and more. These guidelines are designed to empower developers rather than impose control, equipping them with vital tools to enhance the deployment and performance of algorithms. They acknowledge the distinct strengths of each component within a Multi-Scale HPC system. Using multidimensional analysis and a comprehensive strategy, we can establish measurable sustainability metrics, observing all elements of the system. Additionally, these guidelines clarify the configurations for data and application programming, such as setting up a Kubernetes infrastructure that aligns with specific metrics and limitations. This approach helps characterize both multi-HPC systems and applications, ensuring a sufficient standard of performance accuracy.

Ongoing efforts focus on meeting the rapidly growing data needs of fields such as quantum computing and large-scale AI. This involves using Multi-Scale HPC systems with specialized modules—such as those for quantum simulators, rather than real quantum computers. These developments address not only data volume but also the management of large datasets, storage issues, speed enhancements, and data integration complexities. As data grows, AI algorithms tend to improve in accuracy, quantum computing can better maintain coherence and reduce errors, and scientific fields benefit from increased scalability, leading to better simulation results and more precise calculations. Clearly addressing these factors is essential, as they directly impact sustainability metrics.

\section{Conclusion}
\label{CON}

The results presented in this paper showcase their relevance across a range of workloads, primarily low-interaction types similar to those encountered in scientific applications, and high-interaction varieties, such as those found in AI applications. Subsequently, metrics were proposed from a multidimensional perspective to facilitate a comprehensive understanding of the system. These guidelines aid in identifying behavioral patterns by examining sustainability metrics and in characterizing a ratio that outlines Multi-Scale HPC systems, thereby providing insights for behavior prediction and configuration enhancement.

The Characteristic Ratio ($\text{CR}$) framework deliberately excludes a utility-based \emph{Error Tolerance Cost} for two main reasons. First, such a cost (e.g., $X$ lost per 1$\%$ decrease in accuracy) is subjective, highly dependent on the application owner, and would undermine the $\text{CR}$'s goal of being an objective metric based solely on physical and architectural data. Second, this cost is already implicitly addressed by the denominator through the Required Accuracy ($\text{RA}$) constraint. Observing the last Figure \ref{fig:EnergyAccuracy}, it demonstrates that the system must meet the $\text{RA}$ requirement to achieve a meaningful score; if a module fails to reach this minimum accuracy, its $\text{CR}$ score will be disproportionately low, making it unsuitable for allocation regardless of the module's energy efficiency. This is crucial for our proposal because it ensures that the sustainability guidelines derived from the $\text{CR}$ are universally applicable across financial models, providing a hardware-centric, physically grounded guide for optimal resource selection \cite{b21c} \cite{b21d}.

The proposed metrics and guidelines indicate that monitoring loses effectiveness once the application's key features are identified, such as in well-known applications or pre-profiled containerized environments. Instead, it is essential to develop policies and strategies that promote execution and optimization within modular systems, especially in modern Multi-Scale HPC systems. These strategies aim to enhance configurations involving orchestrators, schedulers, and deployers to boost computational efficiency \cite{b22}. For example, Kubernetes cluster setups can be recommended for large-scale systems, directing traffic to specific modules based on workload and energy efficiency goals. However, because workloads are diverse and complex, and due to the modular structure of Multi-Scale HPC systems, assessing individual metrics alone is inadequate. Our advice emphasizes that effective characterization depends on linking various metrics to the specific analysis context, allowing accurate profiling and characterization of system components and execution types.

In conclusion, our study emphasizes the urgent need for effective Resource Allocation and Sustainability Guidelines in Multi-Scale Hybrid HPC Architectures. We propose the Characteristic Ratio ($\text{CR}$), a detailed, multi-dimensional metric that goes beyond basic energy-per-operation measurements. The $\text{CR}$ assesses trade-offs among architectural performance, utilization, and energy efficiency, while also considering Required accuracy ($\text{RA}$), linking sustainability to workload quality. Unlike subjective or cost-based models, our framework provides a physically justified method for identifying the optimal operating point, thereby maximizing computational efficiency and reducing environmental impact. These guidelines enable HPC administrators to make informed deployment decisions, fostering resource allocation that balances speed with sustainable, high-quality performance across the Computing Continuum.

\
\section*{Acknowledgments}

 The authors thank L. A. Torres, S. Gelvez, F. Mejia and P. Rojas for data on their experiences on various platforms, particularly the Grid5000 platform supported by INRIA and its scientific interest group, including CNRS, RENATER, multiple universities and organizations, and the SC3UIS Center \footnote{\url{https://www.sc3.uis.edu.co}}.

\bibliographystyle{IEEEtran}
\bibliography{references}


\end{document}